\documentclass[twocolumn,amsmath,amssymb,floatfix,nature,shownopacs,footinbib,superscriptaddress]{revtex4-2}
\usepackage{amsmath,amssymb,natbib,bm,url,epsf,color,siunitx,float}
\usepackage[noend]{algorithm}
\usepackage[noend]{algpseudocode}
\usepackage[pdftex]{graphicx}

\renewcommand{\Phi}{\varPhi}

\usepackage{xcolor}

\begin{document}

\title{Critical mobility in policy making for epidemic containment}

\author{Jes\'us A. Moreno L\'opez}\affiliation{Instituto de F\'{\i}sica Interdisciplinar y Sistemas Complejos IFISC (CSIC-UIB), Campus UIB, 07122 Palma de Mallorca, Spain}

\author{Sandro Meloni}\affiliation{Instituto de F\'{\i}sica Interdisciplinar y Sistemas Complejos IFISC (CSIC-UIB), Campus UIB, 07122 Palma de Mallorca, Spain}

\author{Jos\'e J. Ramasco}\affiliation{Instituto de F\'{\i}sica Interdisciplinar y Sistemas Complejos IFISC (CSIC-UIB), Campus UIB, 07122 Palma de Mallorca, Spain}

\setlength\parindent{0pt}

\begin{abstract}
When considering airborne epidemic spreading in social systems, a natural connection arises between mobility and epidemic contacts. As individuals travel, possibilities to encounter new people either at the final destination or during the transportation process appear. Such contacts can lead to new contagion events. In fact, mobility has been a crucial target for early non-pharmaceutical containment measures against the recent COVID-19 pandemic, with a degree of intensity ranging from public transportation line closures to regional, city or even home confinements. Nonetheless, quantitative knowledge on the relationship between mobility-contagions and, consequently, on the efficiency of containment measures remains elusive. Here we introduce an agent-based model with a simple interaction between mobility and contacts. Despite its simplicity our model shows the emergence of a critical mobility level, inducing major outbreaks when surpassed. We explore the interplay between mobility restrictions and the infection in recent intervention policies seen across many countries, and how interventions in the form of closures triggered by incidence rates can guide the epidemic into an oscillatory regime with recurrent waves. We consider how the different interventions impact societal well-being, the economy and the population. Finally, we propose a mitigation framework based on the critical nature of mobility in an epidemic, able to suppress incidence and oscillations at will, preventing extreme incidence peaks with potential to saturate health care resources.
\end{abstract}

\maketitle

The importance of human mobility in shaping the spreading of infectious diseases is one of the pillars of Computational Epidemiology \cite{salathe2012,barbosa2018,grantz2020,buckee2020,oliver2021}. Along the years, hundreds of works confirmed the paramount role of mobility \cite{flahault1992,grais2003,brownstein2006,hufnagel2004,colizza2006,colizza2007,balcan2010} -- at all scales: from urban commuting \cite{charaudeau2014,moss2019,massaro2019,changruenngam2020,badr2020,brizuela2021} to intercontinental air travel \cite{hufnagel2004,tatem2012,zhang2017}  -- in driving diseases diffusion. For this reason, containment policies based on non-pharmaceutical interventions often involve a strong reduction in human mobility and different levels of confinements with the aim of reducing contacts within and between populations \cite{bajardi2011,poletto2014,fang2020,chinazzi2020}.  

\bigskip

A paradigmatic example of the application of these strategies have been the measures to fight against the COVID-19 pandemic. Forced by the lack of pharmaceutical solutions and the availability of vaccines, most countries around the world had to rely on social-distancing measures and mobility restrictions to slow-down community transmission of SARS-CoV-2 and limit the collapse of healthcare systems \cite{perra2021}. Although countries adopted different types of interventions \cite{haug2020}, all aim at reducing social mixing by limiting mobility, creating social clusters where cases can be detected and isolated. For almost all the strategies, policy-makers linked the strength and duration of interventions to the epidemiological situation \cite{haug2020,ecdcmovement} --i.e. imposing thresholds on different parameters such as the effective reproduction number $R_{\textit{eff}}$, the disease incidence per 100.000 individuals or the number of ICU patients. Those policies that, in most cases, implied home confinement \cite{gatto2020,didomenico2020}, curfews \cite{didomenico2021} and mobility reductions of more than $80 \%$ \cite{schlosser2020,gibbs2020,kraemer2020} with respect to normal periods, eventually led to a consistent reduction of COVID-19 incidence and transmission rates \cite{maier2020,salje2020,flaxman2020,cowling2020}, with data directly linking mobility reductions and decreased community transmission \cite{unwin2020,vollmer2020,aguilar2023,kissler2020}. 

\bigskip

However, lifting limitations led to a rapid resurgence of incidence cases and transmission \cite{bertuzzo2020,lemey2021,mellan2020} with many countries having experienced a $5^{th}$ or even a $6^{th}$ wave of the epidemic \cite{whosituation}. This oscillatory behavior, as we will demonstrate, can be attributed to delays {and inaccuracies} intrinsic to epidemiological measures. The effective reproduction number $R_{\textit{eff}}$, for example, estimates the average number of secondary cases generated by an infected individual. However, its measurement is affected by biological, social and technological lags -- e.g. the disease incubation period or delays in testing and processing. Thus, $R_{\textit{eff}}(t)$ measured today accounts for infections occurred in the previous 2 or 3 weeks. Similar delays are inherent to other parameters as well. E.g. the number of ICU patients or deaths suffer from even larger delays since several weeks can pass from infection to hospitalization or death. Combined together these delays create a shift between the time of infection and the evaluation of the epidemiological situation, that is, measurements represent an epidemiological scenario up to 3 or 4 weeks in the past, making hard a real-time control of the transmission. In a recent work, Nouvellet et al. \cite{nouvellet2021} proposed a statistical framework to directly link human mobility in one area at time $t$ with the effective reproduction number $R_{\textit{eff}}(t)$ in that area. Their results, relative to COVID-19 infections in $52$ countries, demonstrate an excellent agreement between the two measures, especially for the first wave of the epidemic, when individual immunity was extremely low and other preventive measures like face-masks and closure of indoor spaces were absent. These results, along with providing a methodology for estimating $R_{\textit{eff}}(t)$ from mobility data, also have a more subtle implication: the existence of a critical mobility threshold below which $R_{\textit{eff}}(t)<1$ allowing for a direct control of community transmission. 

\bigskip

Here we develop a simple, yet meaningful, modeling framework able to grasp the connection between local mobility and individual contacts. In this way we are able, not only to confirm the existence of the critical mobility range needed to contain community-level spreading hypothesised empirically, but also to estimate its value for different immunity levels. Moreover, building on the link between mobility range and the effective reproduction number, we revisit intervention principles employed in the recent pandemic and propose an epidemic containment strategy where mobility range is used to assess the epidemiological situation and guide interventions, reducing the delay between epidemiological measurements and interventions. Although in practice an accurate estimation of the critical mobility level may not be available, we develop an heuristic procedure based on the knowledge gained from this model to approximate this range and gain its benefits with minimal information. Finally, reviewing the different strategies improves our understanding of the consequences of each intervention and can help make informed decisions to minimise healthcare system's overload while controlling stress in both the economy and the population.

\section*{Methods}

\subsection*{Minimal data layer}

We use a substrate of data to make the model minimally realistic, but without losing generality. Any population and mobility data can be used or even a brand new one can be generated. In this context, we have employed the population distribution of the city of Madrid together with cell phone data, dividing the province around the city in 293 polygons. These polygons correspond to mobility areas, with all guarantees of preserving privacy, as described by Spain’s National Statistics Institute (INE) {\cite{est_exp_INE}}. In each polygon, we count with an estimation of the resident population based on census. {The data is comprised of aggregated trips between residence and workplace areas, having been assigned to every user according to the most common location from 01:00 to 6:00 AM and from 10:00 AM to 08:00 PM respectively.}


\subsection*{Model}
\label{sec:intro}

\begin{figure}[t!]
\centering
\includegraphics[width=\linewidth]{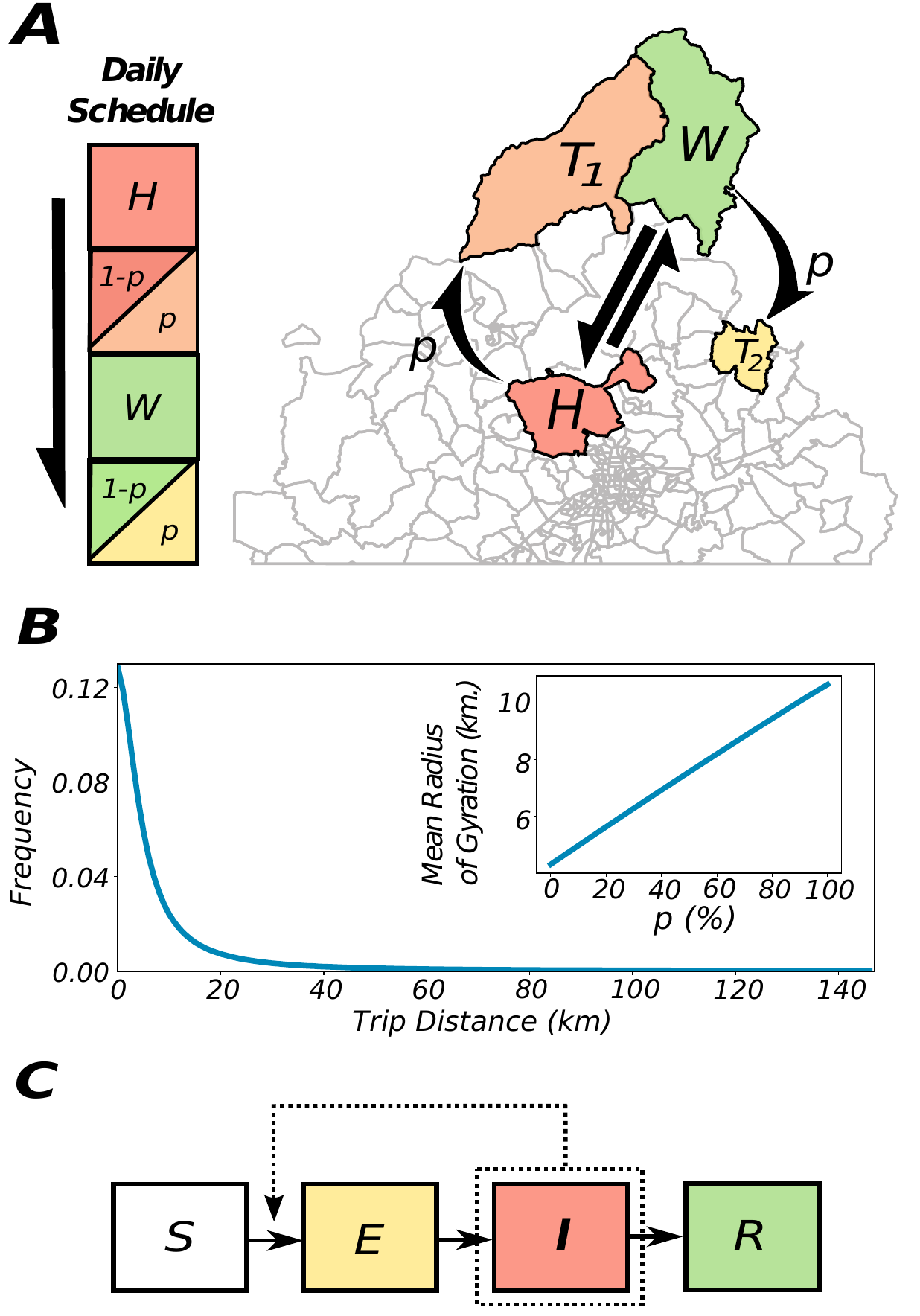}
\caption{Summary of model. \textbf{A} Sketch of zones where agents can be seen during the day. \textit{H} and \textit{W} represent house and workplace respectively, with fixed mobility between them represented in black arrows. $\textit{T}_1$ and $\textit{T}_2$ represent extra trips that may occur with a probability \textit{p}, represented with curved white arrows. \textbf{B} Trip distance statistics from our model employing equation \eqref{eq:cauchy} and relationship between our mobility metric $R_g$ and the trip probability \textit{p}. \textbf{C} Compartmental model. Transition rates and other parameters are listed in Table \ref{tab:rates}. The compartments consist of the following states: susceptible $S$, exposed $E$, infected $I$ and recovered $R$.}
\label{fig:model}
\end{figure}

We developed an agent based model which has $10^6$ agents moving and producing social contacts {via visiting} subdivisions of the Madrid region in Spain. Agents daily commute between a home (\textbf{\textit{H}}) and workplace (\textbf{\textit{W}}). Moreover, along with the home-work commute, they can also perform an additional trip from both home and workplace with a probability {\textit{p}}, as shown in Fig. \ref{fig:model}A. The extra trips given by {\textit{p}} are decided by selecting areas at a distance \textit{d} sampled from a Cauchy distribution bounded by the region of study (Fig. \ref{fig:model}B and see next section). The trip destination is chosen from a population weighted sampling of all regions at distance \textit{d}. {The distribution and population weighted method of generating extra mobility patterns is consistent with the literature \cite{understanding,universal,zipf1946} and can be substituted by real data if such data happened to be available}. 

\bigskip

Spreading of the disease is simulated via a mass-action principle applied to the occupants of each subdivision of the region. Fig. \ref{fig:model}C shows the compartmental model used to represent the different stages of infection, with rates and generation times similar to the recent COVID-19 pandemic{, although any infectious process with non-zero incubation period will qualitatively display the same phenomenology}. The compartments are comprised of an infectious state $I$ and non-infectious states $S$ for susceptible, $E$ for exposed and $R$ for recovered. All simulations start with the introduction of a seed of 100 infected, i.e., an incidence at $t=0$ of 10 cases per 100 000 individuals.

\subsubsection*{Implementation of Mobility}

We assume that the mobility home-work (H-W) is always present (black arrows in Fig. \ref{fig:model}A). Additional {contacts and mobility, called Trips'' ($T$),} can be included from any of the two locations $H$ and $W$ with a {trip} probability $p$. By tuning the parameter $p$, we can thus pass from a situation with a basal mobility to other {situations} with a maximum of 4 social trips {per agent}. For convenience, we can refer to these 2 periods of the day related to Home+Trip 1 and Workplace+Trip 2 as ``morning'' and ``afternoon''. So a standard day is formed at least by a trip from home to work in the morning plus the return in the afternoon, with one potential extra trip in each time period. The distance {to each destination of Trip 1 ($T_1$) and Trip 2 ($T_2$)} is sampled from a discrete version of a Cauchy distribution (Fig. \ref{fig:model}B):
 \begin{equation}
 \centering
\label{eq:cauchy}
P(d) = \frac{1}{Z \left[1+\left(\frac{d}{\gamma}\right)^{2}\right]},
\end{equation}

where $Z$ is a normalization constant and $\gamma = 5 km$ so that 99\% of the distances sampled fall under 147 km, the maximum distance that can be traveled within the Madrid region. 

\bigskip

We choose this particular form as it represents a simplification of previously observed power laws in mobility \cite{understanding,universal} without encountering divergence at low distances. Due to the social nature of mobility, once the distance $d$ is determined, the destination of that trip will be sampled from the locations at distance $d$ to the origin, using their population as weights for the sampling {\cite{zipf1946}}.

\subsubsection*{Epidemic spreading}

We use a compartmental model with structure sketched in Fig. \ref{fig:model}C {and} transition parameters listed in Table \ref{tab:rates}. Most individuals start in the susceptible $S$ compartment, and {contacts with infectious agents may lead them to the exposed $E$ stage}. Exposed individuals are developing the disease but are not yet contagious. {The compartments correspond to an SEIR model, the simplest compartmental model to display a non-zero mode in the generation time distribution \cite{vazquez2020}, relevant for the interplay between generation times, detection and mobility, plus being pertinent to most epidemic-prone diseases. 

\newpage

The transition rates between compartments are inspired by the current COVID-19 pandemic (see the parameter sources listed in Table \ref{tab:rates}), with no intention to replicate this disease with major accuracy, desiring only to obtain realistic timescales for the different infection stages.}

\bigskip

\begin{table}[h]
\begin{tabular}{cr}
\hline \multicolumn{1}{|c|}{Transition}          & \multicolumn{1}{c|}{Rate} \\ \hline
\multicolumn{1}{|c|}{$E \rightarrow I$} & \multicolumn{1}{r|}{$(5.2 \hspace{0.3em} t.s.)^{-1}$}     \\ \hline
\multicolumn{1}{|c|}{$I \rightarrow R$} & \multicolumn{1}{r|}{$(7 \hspace{0.3em} t.s.)^{-1}$}       \\ \hline
\end{tabular}
\caption{Transition rates of the epidemic compartmental model. Values have been selected to be compatible with the timescales of COVID-19 infections \cite{moreno}. $t.s.$=timesteps.}
\label{tab:rates}
\end{table}


We assume that individuals have an average of $k$ contacts {per location}. If we call the infectiousness of the disease per contact $c$ and the probability of infection {upon visiting a location} $P$, for every susceptible in a population of \textit{N} individuals with $I$ infected follows a mass action principle in the form:

\begin{equation}
P_{i} = \beta \frac{I}{N-1}, 
\label{eq:gen_mass_action}
\end{equation}

where the infectivity is $\beta = c\, k$. {As this is true per location, a higher mobility implies more locations per agent, thus an augmented number of contacts between individuals in our population}. The baseline contact level, {always present}, is given by interactions in the Home and Workplace areas. The visit to other places, $T_1$ or $T_2$, amplify the number and variety of potential contacts. {To ensure no contacts are repeated when there is no extra social mobility associated to it, a pair of agents with no extra-trips will not interact twice at home or the workplace.}

\subsubsection*{Characterizing population mobility levels}

{The agent-based nature of our simulations allows us to directly explore and employ methodology tried and tested for recent data driven studies of human mobility.} To have a quantitative metric on the level of mobility within the population, we use the so-called radius of gyration $R_g$ \cite{understanding,Hernando}. For an individual $i$, we can define a radius of gyration, $R_{g}$ based on movements from the residence location as

\begin{align}
\centering
R_{g, i}=\sqrt{\frac{1}{N_i} \sum_{j=1}^{N_t}\left|\left|\mathbf{r}_{j,i}-\mathbf{r}^i_{H}\right|\right|^{2}}
\end{align}

where $\mathbf{r}_{j,i}$ is the vector marking the centroid of an area $j$ visited by agent $i$, $\mathbf{r}^i_{H}$ is the vector pointing to the centroid of the area of residence ($H$) of the agent and the index $j$ runs over the $N_i$ trips of the agent. Increasing $p$ will increase the number of trips performed by an agent, {thus inducing} a larger radius of gyration.  

\newpage

To characterise the whole mobility level in the population, which we will name $R_g$, we calculate the mean of the distribution of all $R_{g,i}$ of all the agents so

\begin{equation}
\centering
R_g = \langle R_{g,i} \rangle
\end{equation}

By taking the mean of the distribution, $R_{g}$, provides a completely smooth, linear and unambiguous relationship with $p$ as can be seen in the inset in Fig. \ref{fig:model}B. It is important to stress that $R_g$ is an observable that can be measured both in the model and in the empirical data -e.g. out of mobile phone records-, while the control parameter $p$ is an abstraction that refers only to our model. {Thus, to measure results, we use $R_g$ to quantify mobility from now onward}.

\section*{Results}

\subsubsection*{\textcolor{black}{Epidemic threshold connection to mobility}}

\begin{figure}[t]
\centering
\includegraphics[trim=15 10 50 30, clip, width=\linewidth]{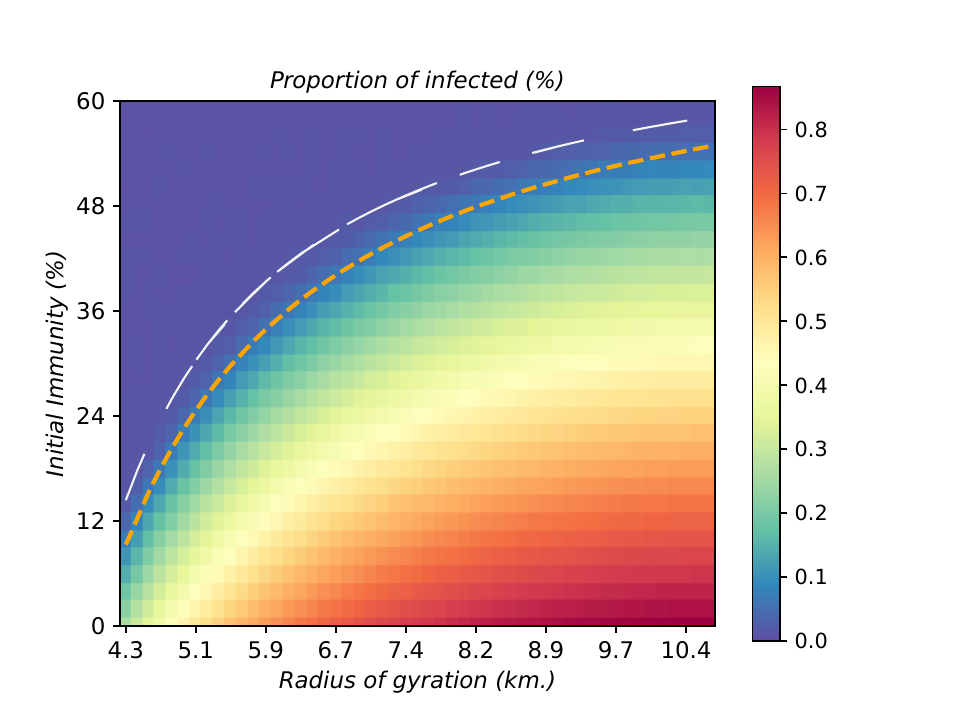}
\caption{Epidemic size, i.e., the fraction of infected individuals at the end of the simulation, as a function of the population mobility $R_g$ and the initial immunity levels for infectivity $\beta = 0.1$. The dashed white line shows eq. (\ref{eq:fit}) performed to points with epidemic size between $1.3\%$ and $2.5\%$, used to estimate the critical mobility threshold $R_{g,c}$ based on immunity. Equivalently, the dashed orange line is another fit performed to mark the boundary below which mobility induces at least a 5\% outbreak.}
\label{fig:imm_map}
\end{figure}

As a direct consequence of the relation introduced between contacts and mobility, our model shows an epidemic threshold that depends on both the immunisation and the population mobility level via its descriptive metric $R_g$. Note that this has been a feature hypothesised out of empirical data in the literature \cite{nouvellet2021,Hernando}. Fig. \ref{fig:imm_map} exposing how much mobility is necessary to spark an epidemic with a seed of infected under different initial proportions of immune population. The dependent variable displayed in the figure is the final size, understood as the fraction of population infected at the end of our simulations. 

\bigskip

A mobility boundary dictated by the immunity level in the population is observed{, inducing 2 phases; one with outbreaks and one without any epidemic activity}. The boundary between large and small outbreaks is artificially set at a final proportion of infected population between $1.3$\% and $2.5$\%. With mobility below such boundary (left of the dashed line), only fewer than 1\% of the population gets infected with the disease, while higher values of $R_g$ (right of the line) lead to major outbreaks. The observed transition allows us to define a series of boundary points in which to perform a fit and obtain an estimation for $R_{g,c}$, the critical mobility threshold estimated for each level of immunity in the population. For practical purposes, we fit $R_{g,c}$ with:

\begin{align}
\label{eq:fit}
 R_{g,c} (km) \approx 2.56 - \frac{1.03}{x-0.71}
\end{align}

where $x$ is the fraction of immune individuals in the population. This simple equation allows us to interpolate the critical curve as a function of $R_g$. 

\newpage

Our model with other parameters and other more realistic models for COVID-19 as the one of Ref. \cite{didomenico2020} (see Supplementary Material, SM, Fig. S1D) also produce a similar critical curve for the relation between mobility and affected population for an intermediate range of infectivity values, especially for those in which the final fraction of infected lays between $10\%-90\%$, as in this range containment interventions on mobility can have an effect. On the extremes, if $\beta$ is very large, the critical boundary will shift upwards in Figure \ref{fig:imm_map}, making nearly all infections happen at the minimum mobility, thus leaving a minimal chance for improvement with interventions (see SM, $\beta = 0.25$ in Fig. S2). Conversely, with a very small infectivity, the critical boundary will shift downwards and thus herd immunity will be located at a very small fraction of immune, limiting the length and epidemic size of our simulations and leaving us, yet again, a small margin to evaluate the effectiveness of measures (see SM, $\beta = 0.05$ in Fig. S2).

\bigskip

{The fact that the critical immunity at the level of minimal mobility is non-zero induces an uncontrollable first outbreak. The uncontrollable nature of the outbreak can be understood from Fig. \ref{fig:imm_map}; starting at the origin of the figure with no immunity and no extra-mobility, which already entails a noticeable outbreak, the only possible strategy is to follow a vertical trajectory from the origin, building immunity while at $p=0$ to get above $R_{g,c}$, at which point mobility can be leveraged around the critical curve}.

\subsubsection*{\textcolor{black}{Interventions}}

Two possible types of non-pharmaceutical interventions are implemented in our model. All of them aim to control spiking incidences via a reduction of $p$ and, consequently, of the contacts between the individuals. Simulations start at $p=0$ and try to drive the system to $p=1$ with the least epidemiological impact possible. In the first intervention procedure, inspired by the COVID traffic light system introduced in many European countries, the mobility reductions are triggered by exceeding the established incidence thresholds \cite{haug2020}. The second type of intervention takes $R_g$ as the main reference variable, trying to keep mobility under control and to follow the curve of $R_{g,c}$ in Fig. \ref{fig:imm_map}. This can be achieved by either estimating the value of $R_{g,c}$ via data and epidemiological forecasting or via tentative approximation of the $R_{g,c}$ curve due to its monotonously increasing nature with immunity. Various epidemiological and mobility indicators help us compare each of the interventions.

\bigskip

Besides non-pharmaceutical interventions, we can also consider two scenarios regarding population immunisation: i) a baseline scenario without vaccination in which only infected individuals can enter the $R$ compartment, and ii) mimicking a vaccination campaign {where} susceptible individuals $S$ pass directly to the removed compartment $R$ at a constant rate each time-step. We show results for the no vaccination scenario as a progressive vaccination campaign yields no qualitative difference {in the interplay between mobility and the epidemic} and constitutes only a narrowing of the simulation window before herd immunity at $p_c \to 1$ is attained (fig. S3 in SM).

\bigskip

In simulations, we can find a numerical variability between realisations. For example, some of them may not develop an outbreak. In order to keep this numerical noise under control once fixed the parameters and the intervention protocol, a realisation is deemed as valid if the total proportion of infected individuals surpasses $50\%$ of the population, which is approximately herd immunity in our model at maximum mobility. In this way, we generate an ensemble of model realisations for every scenario. We require that at least $75\%$ of valid realisations to consider an intervention satisfactory and stable, leaving a margin of $25\%$ to avoid to be too strict in suppressing the natural model noise.

\subsubsection*{\textcolor{black}{Incidence threshold-based interventions}}

\begin{figure*}[t]
\includegraphics[width=\linewidth, trim=5 0 0 5]{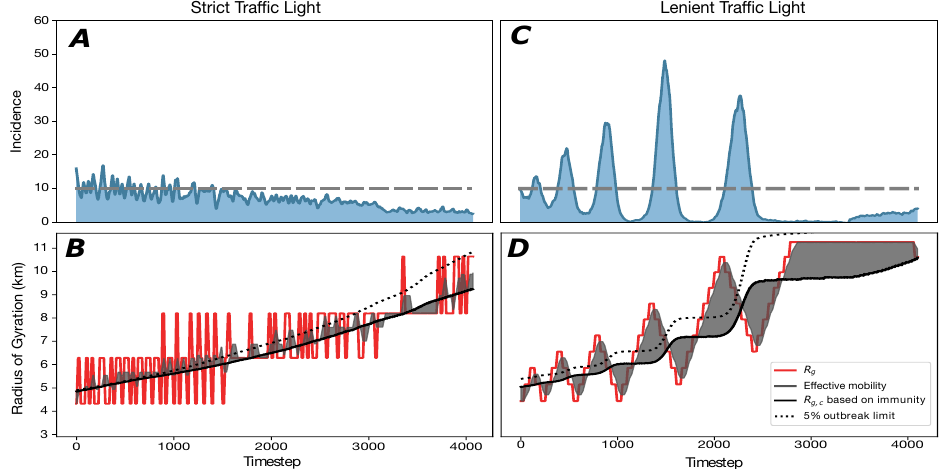}
\caption{Examples of incidence threshold-based interventions for infectivity $\beta = 0.1$. The first uncontrollable epidemic wave is removed due to being completely uncoupled from the dynamics the different interventions induce. (\textbf{A} and \textbf{B}) Incidence per 100 000 inhabitants and mobility curves throughout a realisation of strict traffic light interventions without vaccination. Mobility is adjusted every $T_r=30$ timesteps and adopts the values stated in Table \ref{tab:traffic_light} immediately. The grey dashed line in \textbf{A} and \textbf{C} represents an incidence rate of 10 daily cases per 100 000 individuals. (\textbf{C} and \textbf{D}) Epidemic and mobility curves throughout a realisation of lenient traffic light interventions without vaccination. Mobility aims to adopt the values stated in Table \ref{tab:traffic_light} progressively, in changes of $\Delta p = 10\%$ revised every $T_r=60$ timesteps. \textbf{B} and \textbf{D} include the 5\% outbreak mobility limit plus a 90 timestep moving average of $R_g$ as a measure of an effective mobility.}
\label{fig:main_res}
\end{figure*}

In this type of interventions, the main variable to monitor is the accumulated incidence per $100,000$ inhabitants during the last 14 days, $CI_{14}$. The ideal objective is to have the epidemic under control whilst driving the mobility as swiftly as possible towards $p = 1$, mimicking the lifting of control measures by the authorities. 

\newpage

The value of $p$ allowed is revised every $T_r$ time steps and adjusted following table \ref{tab:traffic_light} in correspondence to the cumulative incidence. The selected values are arbitrary and another set can be selected to work with. {In an effort to make the simplest assumptions about this procedure, we choose equally-spaced intervals for the mobility parameter $p$ and incidence intervals similar to those seen in the recent COVID-19 pandemic \cite{actuaciones,estrategia}. Due to the segmented, start-stop, nature of the intervention, we will interchangeably refer to this intervention as ``traffic lights''. We define a ``strict traffic light'' intervention as an incidence threshold-based intervention where mobility values are adjusted immediately whenever a revision is held.}

\begin{table}[h]
\begin{tabular}{|c|c|}
\hline
$CI_{14}$            & $p$   \\ \hline
{[}0,50{)}     & 1   \\ \hline
{[}50,100)        & 0.6 \\ \hline
{[}100,150)       & 0.3 \\ \hline
{[}150,$\infty$) & 0   \\ \hline
\end{tabular}
\caption{Mobility states, i.e., values for the trip probability $p$, utilised to control mobility for each 14 day accumulated incidence rate $CI_{14}$ in our incidence threshold-based interventions.}
\label{tab:traffic_light}
\end{table}

{In a real case scenario, enforcing drastic and strict restrictions may be unfeasible in a practical sense and might very well also entail enormous impacts to societal well-being and economic stability. Thus exploring an inaccurate adoption of measures is relevant. In consequence, we introduce a less stringent intervention, characterised by higher revision times $T_r$ and by limiting the magnitude of closures and re-openings, even if it means reaching the intended mobility state late. In this intervention, the maximum allowed adjustment in the parameter $p$ is described by $\Delta p = (p_{\text{new}} - p_{\text{old}}) / p_{\text{old}}$. The aforementioned procedure defines a ``lenient traffic light'' intervention}.

\bigskip

Fig. \ref{fig:main_res} exemplifies how {incidence based interventions induce oscillations in both the epidemic and mobility. More so,} drifting away from a strict approach to this system not only significantly increases the period of oscillations but also their amplitude. {This can be attributed to extended times of $R_g$ both above and below the critical mobility level $R_{g,c}$ as shown in red in Fig. \ref{fig:main_res}B and D, causing higher peaks and deeper valleys as shown in figure Fig. \ref{fig:main_res}A and C. The strict approach, even when the traffic light values may be distant to the critical mobility curve, approximates the optimum value via fast-oscillations in mobility (Fig. \ref{fig:main_res}B). We can thus measure this behaviour via am ``effective mobility'' defined as the moving average of the $R_g$ curve in a sufficiently large window; which in Fig.\ref{fig:main_res}B stays close to the critical value. In contrast, Fig.\ref{fig:main_res}D shows an effective mobility with a higher deviation from $R_{g,c}$, remaining frequently above the 5\% outbreak mobility limit (orange curve Fig. \ref{fig:imm_map}). 

\newpage

This implies that in a sufficiently small window as to not build epidemiological inertia, strict traffic light interventions manage to approximate the critical mobility level at the cost of frequent closures and re-openings, with the inherent social and economic stress induced.} Metrics to better understand and quantify the results of these interventions are introduced and commented in subsection \textit{Performance differences} below.

\subsubsection*{\textcolor{black}{{$\mathbf{R_g}$-based interventions}}}

\begin{figure*}[t!]
\includegraphics[width=\linewidth, trim=5 5 0 5]{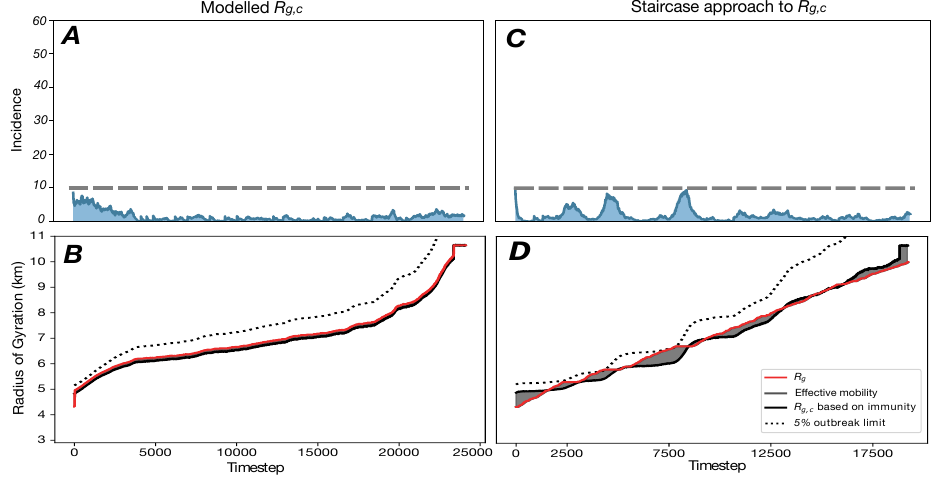}
\caption{Examples of $R_g$-based interventions for infectivity $\beta =0.1$. The first uncontrollable epidemic wave is removed due to being completely uncoupled from the dynamics the different interventions induce. (\textbf{A} and \textbf{B}) Incidence per 100 000 inhab. and mobility curves throughout a realisation of an $R_g$-based intervention with theoretically estimated $R_{g,c}$ and $\epsilon_p = 0.025$ {without vaccination}. Mobility is adjusted every $T_r=30$ timesteps following \ref{eq:rgc-measure}. (\textbf{C} and \textbf{D}) Epidemic and mobility curves throughout a realisation of staircase interventions {without vaccination}. The mobility parameter $p$ is forced to be monotonically increasing through time in steps of $\Delta p = 0.3\%$ every $T_r=30$ timesteps. The grey dashed line in \textbf{A} and \textbf{C} represents an incidence rate of 10 daily cases per 100 000 individuals. \textbf{B} and \textbf{D} include the 5\% outbreak mobility limit plus a 90 timestep moving average of $R_g$ as a measure of an effective mobility. The 90 t.s. average completely overlaps $R_g$ as changes are too slow for them to be distinguishable.}
\label{fig:main_res_2}
\end{figure*}

Imagining that it was possible to {accurately} estimate the mobility threshold $R_{g,c}$ {for any state of our population}, the way to control the epidemic with a minimal intervention would be to conduct the mobility $R_g$ to $R_{g,c}$. Note that as the population gets immunised, either by infection or vaccination, $R_{g,c}$ can only grow, and, therefore, mobility policies would comprise of a gradual, smooth and strictly increasing change of $R_g$ over time, in contrast to the on-and-off dynamics of the traffic light interventions. This is{, from the epidemiological impact perspective,} an optimal type of intervention that we are going to illustrate with our model. Further on we will discuss alternatives for situations in which the curve of $R_{g,c}$ displayed in Fig. \ref{fig:imm_map} is not available. 

\bigskip

We start the simulation by estimating $R_{g,c}$ given the initial immunisation conditions using the fit of Eq. \eqref{eq:fit}. It is important to recall that the curve of $R_{g,c}$ and its fit depend on the epidemic parameters, on the population distribution in space and on the mobility. The fit is only an approximation, since we selected an epidemic size between $1.3$ and $2.5\%$ to establish the boundary between areas with and without outbreaks, which introduces a slight uncertainty and a small potential error. As mentioned in the Methods section, a value of $R_{g,c}$ can be translated into a critical value of $p$, $p_c$, in our model. As in the threshold-based case, we need to establish an intervention revision interval $T_r$. After each $T_r$, the mobility parameter $p$ is increased to match the new estimation of $p_c$.

\bigskip

Unfortunately, calculating $R_{g,c}$ out of empirical data is not a trivial task. An approximated estimation is suggested in recent works such as in Ref. \cite{Hernando} and \cite{nouvellet2021}. For simplicity and to illustrate the power of the critical insight gained thus far, we will later show a heuristic method to remain close to $R_{g,c}$ without delicate modelling and information required. Now, we will use the fit in Eq. \eqref{eq:fit} to obtain an approximation of $R_{g,c}$ (and $p_c$) for our model given the epidemic parameters and a certain initial immunity level in the population. 

\newpage

We plan to stay slightly above the critical level of mobility to lead the epidemic to herd immunity with the minimum incidence possible as vaccination isn't included in this model. Thus, every $T_r$ timesteps, the mobility parameter $p$ is updated to $p_{new}$ following:

\begin{equation}
    p_{new} = \left(p_c\right)_{est.} + \epsilon_p \quad \text{s.t.} \quad R_g \rightarrow \left(R_{g,c}\right)_{est.}+\epsilon_{R}
    \label{eq:rgc-measure}
\end{equation}

\vspace{6pt}

where $(\quad)_{est.}$ indicates an estimation, $\epsilon_p$ and $\epsilon_R$ are small positive integers selected to keep mobility slightly above the critical level and explore the possibility of guiding the simulation to herd immunity via sustaining a minimal incidence. An example of this procedure is shown in Fig. \ref{fig:main_res_2}A and \ref{fig:main_res_2}B. {The intervention manages to maintain a minimal incidence rate whilst monotonically increasing mobility, thus no stress of sudden behavioural changes or difficulties in applying strict changes in short periods apply.} The only relevant incidence spikes are found at the start of the simulations where, due to the pronounced slope of the critical curve for $R_{g,c}$ found from Fig. \ref{fig:imm_map}, any small overshoot in mobility produces a higher incidence outcome. This behaviour of the critical curve at small mobility values combined with an uncontrollable first wave that introduces a high number of cases can cause these small outbursts.

\bigskip

Due to the delayed relationship between epidemiological evolution and mobility/social contacts, the previously mentioned methods for estimating the critical mobility level can only produce results retroactively. This means that the critical mobility level estimated for the time period $\left(t_1,t_1+T\right)$ can only be determined via correlating the corresponding epidemic curves at a posterior time period $\left(t_2,t_2+T\right)$ with $t_2>t_1$ and $T>0$. 

\vspace{20pt}

To estimate the current critical mobility level at any given time with these techniques, a precise epidemiological, demographic and mobility model most be developed to predict the epidemiological outcome of mobility levels at the moment and then use the aforementioned procedures. The complexity and precision these models entail, plus the vast amount of detailed data required to make these predictions accurate, can make this enterprise prohibitive for many governments and institutions, a challenge that becomes particularly difficult when it comes to collecting precise and relevant data on social contacts and patterns.

\newpage

\subsubsection*{\textcolor{black}{Staircase approach to $R_{g,c}$}}

Fortunately, the knowledge gained from analysing the existence and evolution of the critical mobility level can help us design a heuristic to approximate this threshold without needing prior modeling and field knowledge. As we saw in Fig. \ref{fig:main_res}A and B, epidemic thresholds with a sufficiently quick update speed can stabilize the system in a low incidence regime {but a great back on forth in mobility is needed to approximate the critical level.} To avoid this social stress due to fast oscillations in mobility policies and restrictions, we can just attempt to move with the critical mobility level. 

\bigskip

As stated, the monotonous increase of the critical level allows us to correct any overshoot in mobility by maintaining the mobility level until immunity makes the critical level catch up to the mobility. This allows us to design an intervention based on a progressive, yet always increasing re-opening, yielding very small incidence rates with no modeling necessary. This is done by only observing the incidence curves and staying on the safe side of the critical mobility level, only releasing mobility by a small increase if incidence is decreasing and the epidemic is deemed controlled: 

\bigskip

\begin{algorithmic}
\State For every revision of measures:
\\
\If{$CI_{14}<50 \quad \& \quad \Delta I <0$}
    \State $p_{new} = p + \Delta p$;
\Else
     \State $p_{new} = p$;
\EndIf
\end{algorithmic}

\bigskip

Figures \ref{fig:main_res_2}C and \ref{fig:main_res_2}D show the epidemiological outcome and the behaviour of mobility resulting from the aforementioned procedure. Incidence shows only minimal peaks, predominantly below an incidence rate of 10 points, caused by any small overshoot of mobility above the critical mobility level, which we see is easily stabilised by holding mobility in place.

\subsection*{\textcolor{black}{Performance differences}}

We are going to calculate the following epidemic and mobility metrics to each simulation in order quantify the performance of the different intervention protocols:

\begin{itemize}
    \item The average incidence rate.
    \item The proportion of cases occurred above 10 cases per 100 000 individuals, designated as ``surplus incidence'' or `extreme incidence'' in what follows, to focus on possible health care system saturation and also omit large periods of lockdown where the epidemic is not active.
    \item Peak incidence, as another indicator of possible healthcare saturation \cite{didomenico2020}.
    \item Standard deviation of the relative mobility, $\Theta (t) := R_{g}(t) - R_{g,c}(t)$, as a measure of mobility fluctuations relative to the critical level.
    \item {An accumulated measure of the effective mobility above the 5\% outbreak mobility limit as a measure of when, even when oscillating, the intervention is staying a significant time in the outbreak phase of the immunity-mobility diagram.}
\end{itemize}

\begin{figure*}[t!]
\centering
\includegraphics[width=\linewidth, trim = 5 0 0 10]{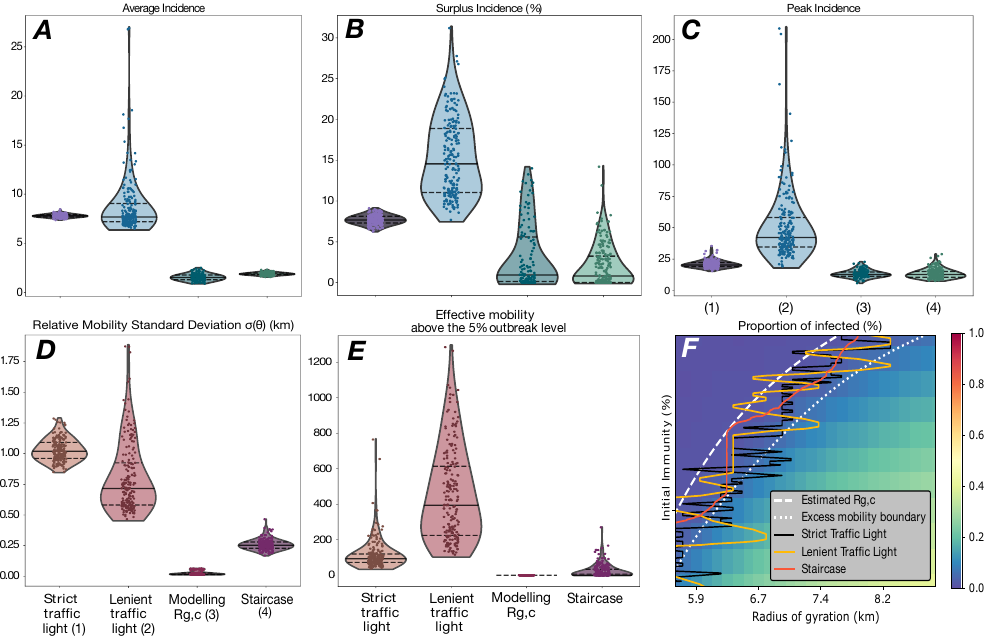}
\caption{Violin and strip plots for statistics of the different metrics and interventions tested. The parameters of each intervention are detailed in figures \ref{fig:main_res} and \ref{fig:main_res_2}. Violin plots include lines to mark median and quartiles of the distribution. The first epidemic wave happens with $p=0$ and thus is uncontrollable. Therefore it is excluded when computing the present statistics to avoid dilution of the results produced in the regions where interventions are acting. The number of simulations for each intervention is 200. \textbf{A} Average Incidence per 100 000 inhab. \textbf{B} Surplus Incidence as a percentage of the total population. \textbf{C} Distribution of Peak Incidences by intervention type. \textbf{D} Deviation of mobility from the optimum, characterised as the standard deviation of the relative mobility $\Theta (t) := R_{g}(t) - R_{g,c}(t)$. {\textbf{E} Cumulative effective mobility (moving average of $R_g$ with a 90t.s. window) above the 5\% outbreak mobility level. \textbf{F} Example trajectories in the phase diagram of Fig. \ref{fig:imm_map} for interventions shown in Fig. \ref{fig:main_res} and \ref{fig:main_res_2}.}}
\label{fig:main_res_3}
\end{figure*}

The various metric results averaged over $200$ realisations of all previously described interventions are summarised in Fig. \ref{fig:main_res_3}. With no vaccination, all simulations arrive stably at a {$>50\%$} epidemic size, as expected from Fig. \ref{fig:imm_map}, independently of the intervention typology.

\bigskip

Both traffic light interventions arrive at a similar median average incidence (Fig. \ref{fig:main_res_3}A), yet a lenient traffic light intervention risks extraordinarily high incidences compared to its counterpart. This is also observed {in the surplus} incidence in Fig. \ref{fig:main_res_3}B. When comparing both figures \ref{fig:main_res_3} A and B we can arrive at the conclusion that a strict implementation of a traffic light systems, as implemented for our model, sustains a stable $\sim$ 10 cases per 100 000 inhabitants until arriving to herd immunity, whilst a lenient implementation of a traffic light system can produce exorbitant oscillations which may be disguised by the average incidence metric but become apparent when contemplating the average extreme incidence metric. This is also aligned with the information appearing in Fig. \ref{fig:main_res_3}C, as all interventions that dampen oscillations and extreme behaviour of an epidemic yield low peak incidence values. These values are centered around $\sim 25$ cases per $100 000$ inhab. as this is the value stemming from the first uncontrollable epidemic wave, setting an upper bound for the worst case scenario in these interventions. A lenient approach to a traffic light system can easily double the peak values or worst, we find peaks up to 8 times as high as the worst case scenario for the other interventions.

\bigskip

Concerning $R_g$ focused interventions, both interventions show extremely low average incidences. Again, we need to observe the values for {surplus} incidence in these cases to understand {if there is any other risk associated to these interventions}. As seen in Fig. \ref{fig:main_res_2}, these interventions cause very mild oscillations which rarely surpass the incidence threshold of 10 cases per 100 000 inhabitants, maintaining the incidence at a nearly minimum value whilst allowing the mobility to steadily increase, causing us to observe minimal average incidences, minimal surplus incidences and also minimal peak incidences. {Curiously enough, even though their median of the average incidence distribution is a whole order of magnitude below the median of threshold-based interventions, it may present outliers which may surpass the former interventions. 

\newpage

This is a fault in the design of these interventions as due to stochasticity, any sudden outburst in incidence, specially frequent at the start of the epidemic where the 5\% outbreak mobility limit is closer to $R_{g,c}$, is incorrigible. More so, the staircase heuristic, even though it's more prone to overshooting as seen in the peak incidence distribution in Fig. \ref{fig:main_res_3}C, seems to have less extreme values of surplus incidence compared to an approach following $R_{g,c}$. This is because it has the ability to detect an extreme event and keep its mobility in stasis, assuring that this outbreak will be eventually controlled}. 

\bigskip

{A similar conclusion can be extracted when observing the mobility above the 5\% outbreak} limit, with staircase interventions naturally showing a higher {chance of potentially overshooting into this region, as modelled $R_{g,c}$ interventions by design will not surpass this bound}.

\section*{Discussion}

The increasing precision and availability of mobility data calls to try new ways in which to monitor and contain epidemics. From individual-based approaches such as app-based digital contact tracing to wider, policy-based approaches such as monitoring commuting flows and general mobility levels, recent studies have been proposing a possible critical mobility level due to the coupling between epidemic outbursts and mobility observed through various sources of data and their derived metrics. In this work, we have introduced a very simple connection between mobility and epidemic contacts, together with a compartmental model with timescales similar to the recent COVID-19 pandemic, and have observed this to be sufficient to not only witness the appearance of a critical mobility level, but also to capture the evolution and behaviour of this critical level throughout an epidemic. 

\newpage

This model has allowed us to {comprehend,} evaluate, and compare different interventions based on their approach towards mobility and the epidemic information available, and has allowed us to {verify} the various effects on the oscillatory behaviour of an epidemic produced by the proposed couplings between mobility and the incidence.

\bigskip

Our agent based model with mobility reduced to a single parameter, the trip probability $p$, allows us perfectly relate epidemic contacts to the empirically measurable observable $R_{g}$. The proposed model also incorporates home-workplace commuting data from the Spanish census, plus known social behaviours such as a probability of trips that decays with the inverse square of the distance. {With these simple assumptions and minimal data, our model is capable of reproducing a phase diagram with a critical mobility level, separating an outbreak phase from a stale phase.} Together with epidemic observables such as the incidence per 100 000 inhabitants, we are able to monitor the state of the system and define interventions to safely conduct the population to herd immunity, {aiming to minimise the oscillation of the system into the outbreak phase}.

\bigskip

Having mainly separated interventions in 2 categories: {threshold-based interventions} and mobility-based interventions, we have been able to understand the principles and consequences behind the use of both kinds of interventions, plus how to better use these interventions to avoid harmful side effects such as saturation of healthcare resources and economic-social stress due to an unstable climate of restrictions and re-openings.

\bigskip

Traffic light interventions are inherently short-sighted and work by applying restrictions and re-openings just by observing the current epidemic state of the system. The fact that the system is forced into a handful of states, in our case 4 (see Table \ref{tab:traffic_light}), depending on the epidemiological state means this system is blind to the key factor in this mobility-driven system; the value of the current critical mobility threshold. Restricting the states of the system means the system requires rapid oscillations between states so that the {effective} mobility{, a key observable to understand the observed phenomena,} is close to the critical threshold in {a sufficiently small window} of time. Additionally, looking only at the incidence to correct the status of your system is {inherently} a sub-optimal approach to interventions as the information provided {inevitably} carries a delay of weeks; the incidence observed at any time is a direct consequence of the events and contacts of a few weeks prior. {These interventions become less accurate with worst case detection, a higher incubation period and a high infectivity of a possible asymptomatic case}. Combining an overlooking of the critical mobility level of the system, the inherent delay and a possible lenient application of the measures will probably drive the system into a highly oscillatory regime of overshooting and undershooting, as seen in \ref{fig:main_res}C and D. Quickness and decisiveness when applying changes in mobility{, plus the social stress they inevitably entail,} are {paramount} if one desires to keep both the average incidence and peaks to a minimum when using a traffic light system, as shown via comparison in Fig. \ref{fig:main_res_3}.

\bigskip

On the other hand, the proposed interventions based on mobility are {fundamentally} about utilising the knowledge and insight gained from observing and understanding the system in terms of its critical mobility level {and outbreak phases}. In the both examples mentioned, {interventions with a modelled $R_{g,c}$} and the staircase heuristic, we either directly set the mobility level to as close to the critical value as possible (assuming it is known) as in the former intervention (Fig. \ref{fig:main_res_2}B) or we try to approximate it by tentatively increasing and sustaining mobility {in reaction to outbreaks} in order to follow its ascent as we did for the latter approach (Fig. \ref{fig:main_res_2}D). These methods seem extremely well behaved, displaying minimal oscillations and very low peak incidences (Figs. \ref{fig:main_res_2}A and C). In essence, the staircase approach to $R_{g,c}$ would also be looking at the $CI_{14}$ threshold, and thus could be subject to the same problems as the traffic light counterparts presented earlier. 

\bigskip

The real difference between the staircase approach and the traffic light system, even though they both look at the epidemic state to decide the next mobility level, is in incorporating the knowledge gained from observing $R_{g,c}$, and most importantly, on the increasing nature of $R_{g,c}$, plus allowing a relaxation of the spectrum of states the mobility can be in. By doing this tentative increase and sustaining of the mobility, the staircase manages to approximate the critical level well enough to not produce large overshoots and undershoots when changing state. This would allow administrations that do not count with precise data and modelling to benefit from the knowledge of the critical mobility threshold. {Whichever the case, utilising the notion of a critical mobility level and an outbreak phase is of the utmost importance in order to minimise epidemic impact.}

\bigskip

Other very important benefits of $R_g$-based interventions are related to the effects interventions and restrictions have on economy and social tissue. Although these interventions do not maximise mobility above $R_{g,c}$ (Fig. \ref{fig:main_res_3}E), they provide less unnecessary closures (lower mobility below $R_{g,c}$, Fig. \ref{fig:main_res_3}F) and less stress from oscillations (Fig. \ref{fig:main_res_3}D). The progressive and monotonous increase in mobility allows for a better adaptation and planning for businesses and individuals, plus it can eliminate the stress and uncertainty sudden and drastic closures may produce on the public. {The most notorious downside of these interventions is their extremely slow rate of achieving herd immunity, achieving simulation times of nearly one order of magnitude higher than traffic light measures. If vaccination is not an option, societies are not able to withstand their economy with closures happening along a very prolonged amount of time. 

\newpage

As we have seen with some countries in the recent COVID-19 pandemic, governments had to choose a trade-off between economic collapse and public health. This downside can be solved with other complementary interventions, as $R_{g,c}$-based measures can buy time and keep the population safe whilst other developments such as vaccines can be developed and put into use, accelerating the journey towards herd immunity.} 

\bigskip

{In practice, a sufficiently accurate model of $R_g$ would provide better results than the one presented in this work. This is due to our choice in modelling as, for simplicity purposes, we keep mobility always a fix amount above $R_{g,c}$. As we have seen, the 5\% outbreak mobility level is closer to $R_{g,c}$ at the start of the epidemic and then leaves more room for error. Most negative effects of this intervention shown in this work are due to not correcting the distance between $R_g$ and $R_{g,c}$ as the epidemic evolves, leaving less distance at the start and increasing mobility respect to $R_{g,c}$ as times advances. In any real scenario, if an outbreak is detected by using this approach, mobility can be held or even lowered if necessary, options not included in our model for simplicity's sake.}


\bigskip

The limitations of our setting and model are evident. Our model assumes mobility as being the only factor in disease spreading and incorporates no information on variable transmissibility, variable behaviour on each stage of the compartmental model, seasonal mobility and changes in social contacts, demography, precise social contact modelling, realistic vaccination campaigns, improved detection and tracking of the incidence, and the problem of how to precisely estimate immunity in the population. Another crucial limitation for a real case application would be that, apart from absolute lockdown, any given mobility policy may have an undetermined reaction by the population. A population may ignore or overreact to a policy depending on their knowledge of the disease, opinion, beliefs and even the amount of time they have been under stress due to containment policies. The recent pandemic has seen some incidence spikes due to pandemic paranoia and lockdown/restriction-induced psychological ill-being in the population, where epidemic risk warnings were relatively or completely ignored due to burnout with the hopes of returning as soon as possible to a normal life regime. Thus a natural progression of this work would be to not only focus on mobility as the single factor to produce this critical phenomena, but also how to incorporate changes in disease transmissibility, include other strains and variants of the disease and changes in average number of contacts, as this would change the parameter $\beta$ without necessarily increasing the mobility level $R_g$. A precise modelling of human behaviour related to complying with policy making and how that affects the observed mobility level and contact patterns would also be necessary to improve results. Note that, although the infectivity depends on other factors, these would only alter the critical boundary in the $R_g$ vs immunity diagram, but not the validity of our analysis and control framework.  {More so, even with our assumptions, mobility and immunity are not the only factors that can cause outbreaks as the current prevalence of the disease in the population plays a great role. Further works must add this third axis to the phase diagram as outbreaks stemming from a given mobility and immunity level will surely be affected by the prevalence in the population.} Possible future works include incorporating better data on contacts and transmission, mobility, incidence and demography in an attempt to produce estimations for the critical mobility level and check via the known methods if a more refined model can accurately predict these critical thresholds. Precise information and observation of the critical mobility threshold would take us a step closer to shifting epidemiological interventions primarily into prevention territory and not mainly into mitigation efforts, as we have seen in the recent COVID-19 pandemic.

\bigskip

Thanks to the knowledge gained from studying the criticality associated to mobility in epidemics and the increasing amount of data available for administrations and research bodies, we hope our study can help to retroactively clarify the role of mobility in the observed phenomena during the recent COVID-19 pandemic together with designing better strategies {for policy making} in the case of future epidemic events; minimising the impact of the disease on the population whilst allowing for a better, smoother transition to normality for both the {economy and society} alike.

\section*{Acknowledgments}

J.A.M.L.’s salary was funded by the Agencia Estatal de Investigación (AEI, MCI, Spain) MCIN/AEI/ 10.13039/501100011033 under grant FPI PRE2019-089734. J.A.M.L., S.M. and J.J.R. also acknowledge funding from the project FACE of CSIC integrated in the platform PTI Salud Global and funded by a contribution of European Recovery Funds. Partial financial support has been received from the Agencia Estatal de Investigación (AEI, MCI, Spain) MCIN/AEI/ 10.13039/501100011033 and Fondo Europeo de Desarrollo Regional (FEDER, UE) under Project APASOS (PID2021-122256NB-C21/C22) and the María de Maeztu Program for units of Excellence in R\&D, grant CEX2021-001164-M).

\section*{Author contributions statement}
J.A.M.L., S.M. and J.J.R. designed research; J.A.M.L., S.M. and J.J.R. performed research; J.A.M.L. analysed data; J.A.M.L., S.M. and J.J.R. wrote the paper; All the authors revised and approved the manuscript.

\section*{Additional information}

\subsection*{Data and code availability}
The mobility data underlying this article is available at the Spanish National Statistics Institute at \url{https://www.ine.es/experimental/movilidad/experimental_em.htm}. The code used to perform simulations is available at \url{https://gitlab.ifisc.uib-csic.es/jeslop/critical-interventions-open}.

\subsection*{Competing interests}
The authors declare no competing interests.

\subsection*{Supplementary Material}
Supplementary Material to this article should be available online.

\bibliographystyle{naturemag}

\bigskip
\bigskip

{\centering \LARGE \textbf{Supplementary Information to Critical mobility in policy making for epidemic containment}}

\section{Complex COVID-19 model \cite{didomenico2020}}

\begin{figure*}[h!]
\includegraphics[width=18cm]{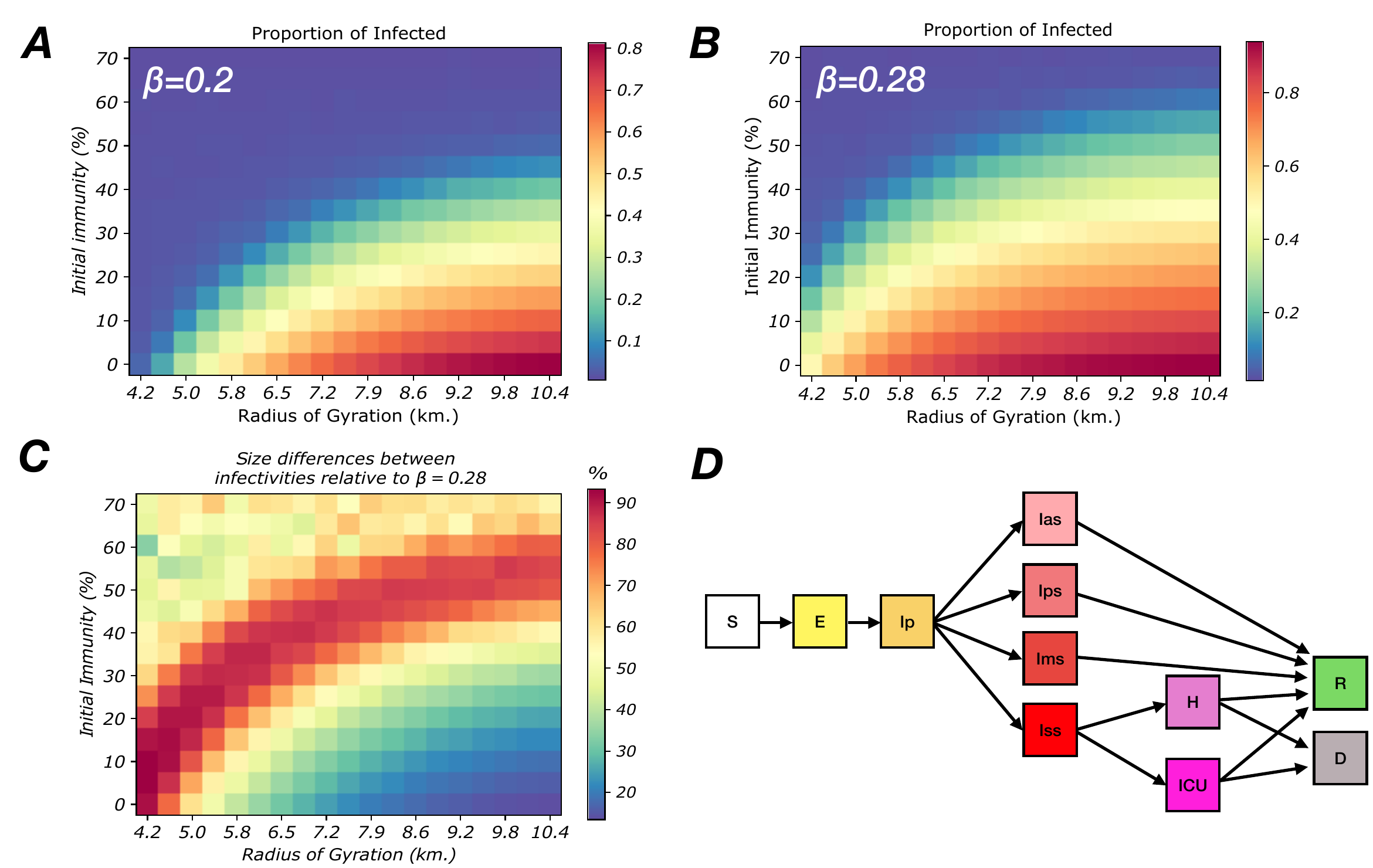}
\caption{\textbf{A}. Epidemic size, i.e., the fraction of infected individuals at the end of the simulation, as a function of the population mobility $R_g$ and the initial immunity levels for infectivity $\beta = 0.2$ in a complex COVID-19 compartmental model \cite{didomenico2020}. \textbf{B}. Epidemic size as a function of the population mobility $R_g$ and the initial immunity levels for infectivity $\beta = 0.28$ in a complex COVID-19 compartmental model \cite{didomenico2020}. \textbf{C}. Size differences between infectivities $\beta = 0.2$ and $\beta = 0.28$ relative to $\beta = 0.28$ for initial immunity levels and population mobility $R_g$. \textbf{D}. Complex compartmental model inspired by \cite{didomenico2020} used in the previous panels. The compartments correspond to: $S$, susceptible; $E$, exposed; $I_p$, infectious in the prodromic phase; $I_a$, asymptomatic infectious; $I_{ps}$, paucysymptomatic infectious; $I_{ms}$, mild symptoms; $I_{ss}$, severe symptoms; $H$, cases admitted into hospitals (excluding intensive care); $ICU$, severe cases in ICU;  $R$, recovered; $D$, deceased. Transition rates can be found in table \ref{tab:supp_vitt_rates}.}
\label{fig:vittoria_dif}
\end{figure*}

To check whether our claims on the critical transition in the immunity-mobility diagram hold for other more complex epidemiological models, we produced other immunity-mobility diagrams with a complex COVID-19 model from \cite{didomenico2020}, increasing the detail even further by excluding severely infected, hospitalised, ICU and deceased individuals from participating in transmission and epidemiological contacts (figure \ref{fig:vittoria_dif}). We can observe how the size and duration of the first epidemic wave will vary between 2 values of $\beta$ for the same model due to the shift of position of the critical mobility boundary curve (red band in figure \ref{fig:vittoria_dif} C).

\begin{table}[h!]
\begin{tabular}{|c|c|}
\hline
Transition                & Rate                       \\ \hline
{$E \rightarrow I_p$} & $(3.7 t.s.)^{-1}$          \\ \hline
$I_p \rightarrow I_{as}$  & $0.35*(1.5 t.s.)^{-1}$     \\ \hline
$I_p \rightarrow I_{ps}$  & $0.65*0.2*(1.5 t.s.)^{-1}$ \\ \hline
$I_p \rightarrow I_{ms}$  & $0.65*0.7*(1.5 t.s.)^{-1}$ \\ \hline
$I_p \rightarrow I_{ss}$  & $0.65*0.1*(1.5 t.s.)^{-1}$ \\ \hline
$I_{as} \rightarrow R$    & $(2.3 t.s.)^{-1}$          \\ \hline
$I_{ps} \rightarrow R$    & $(2.3 t.s.)^{-1}$          \\ \hline
$I_{ms} \rightarrow R$    & $(2.3 t.s.)^{-1}$          \\ \hline
$I_{ss} \rightarrow H$    & $0.76*(2.3 t.s.)^{-1}$     \\ \hline
$I_{ss} \rightarrow ICU$  & $0.24*(2.3 t.s.)^{-1}$     \\ \hline
$H \rightarrow R$         & $0.083 (t.s.)^{-1}$        \\ \hline
$H \rightarrow D$         & $0.0031 (t.s.)^{-1}$       \\ \hline
$ICU \rightarrow R$       & $0.76*(21.1 t.s.)^{-1}$    \\ \hline
$ICU \rightarrow D$       & $0.24*(21.1 t.s.)^{-1}$    \\ \hline
\end{tabular}
\caption{Table with rates for the compartmental model in figure \ref{fig:vittoria_dif}D inspired by \cite{didomenico2020}}
\label{tab:supp_vitt_rates}
\end{table}

\section{Evolution of the critical boundaries}

\begin{figure*}[h!]
\includegraphics[width=18cm]{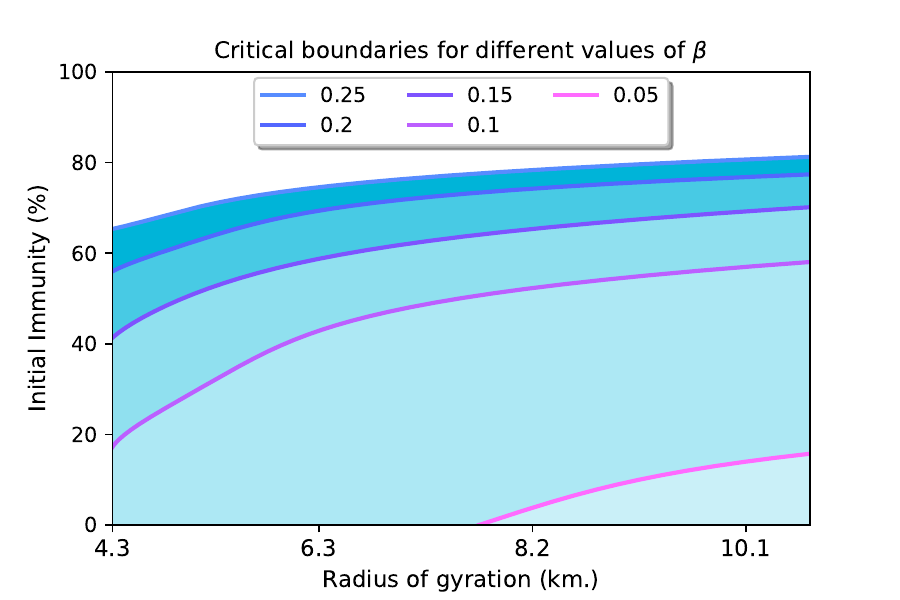}
\caption{Evolution with infectivity $\beta$ of the critical boundary separating the regions of major epidemic outbreaks from stale reactions to seeding in our model.}
\label{fig:crit_bound_beta_simple}
\end{figure*}

In order to demonstrate the importance of the infectivity when it comes to the effectiveness of measures, we have produced figure \ref{fig:crit_bound_beta_simple}, showing the shift. This illustrates the logic and importance behind considering a reasonable range of infectivities.

\section{Effects of vaccination}

\begin{figure*}[h!]
\includegraphics[width=17cm]{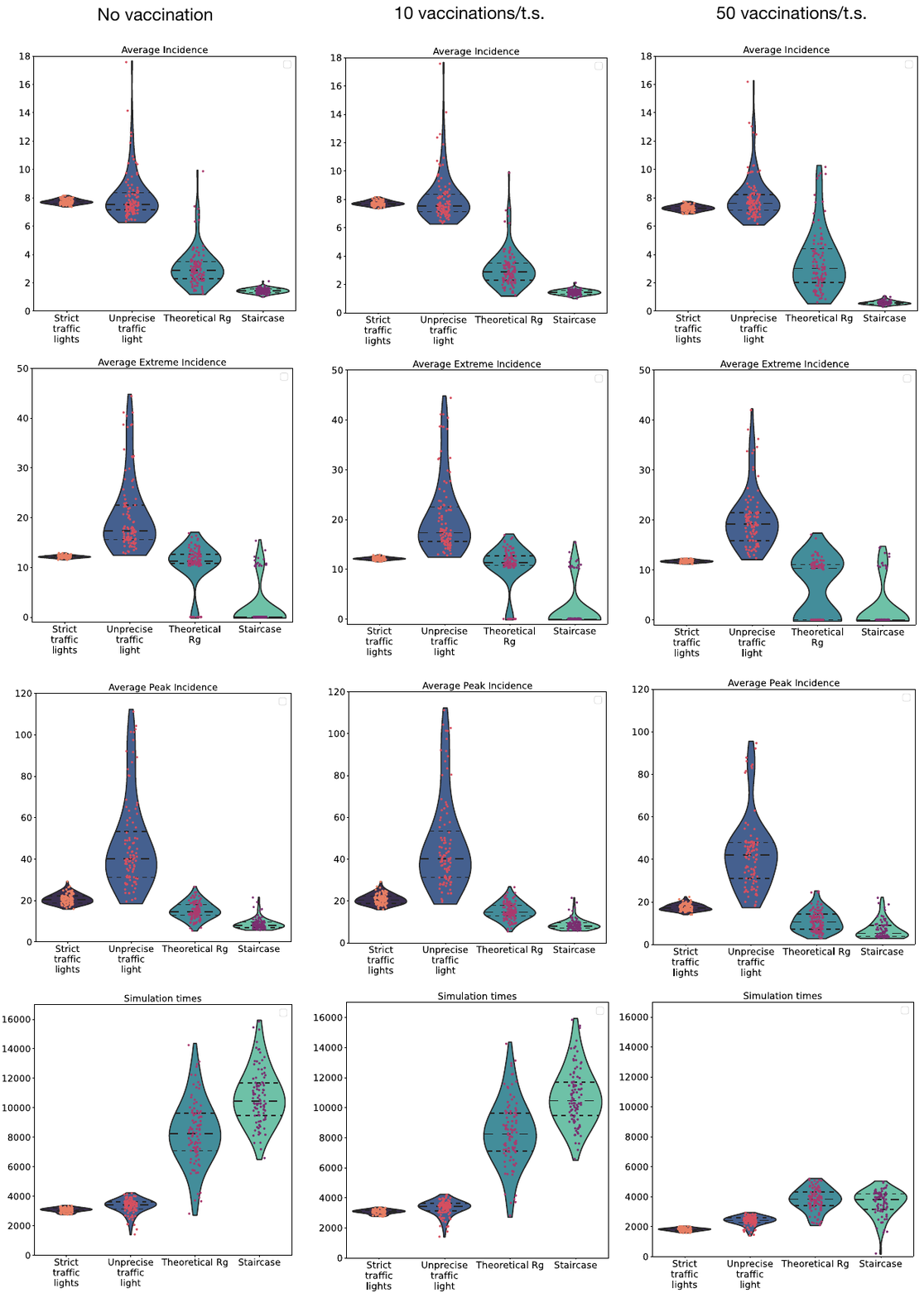}
\caption{Epidemic metrics and simulation times for no vaccination, a vaccination of 10 individuals per timestep and a vaccination of 50 individuals per timestep. The interventions used to generate these statistics are the same as the ones described in the main text.}
\label{fig:stats_vacc}
\end{figure*}

Vaccination introduces immune individuals each timestep, thus reducing the available population for infection and limiting the total length of the simulations. In order to check if vaccination radically changes our results, figure \ref{fig:stats_vacc} compares our epidemic metrics without vaccination to a rate of 10 and 50 vaccinations per timestep. Assuming a herd immunity of around $~60\%$ (Fig. 2), this bounds our total simulation time to $60*10^3$ t.s. for 10 vacc./t.s. and  $12*10^3$ t.s. for 50 vacc./t.s.. Figure \ref{fig:stats_vacc} shows no qualitative change in epidemic metrics between interventions, except the expected shifting of distributions towards weaker outbreaks in the case of a very high vaccination rate, supporting our claims that vaccination qualitatively doesn't affect the results, it only shortens the simulation window.

\newpage

\section{Stability of the traffic light system}

\begin{figure*}[h!]
\includegraphics[width=17cm]{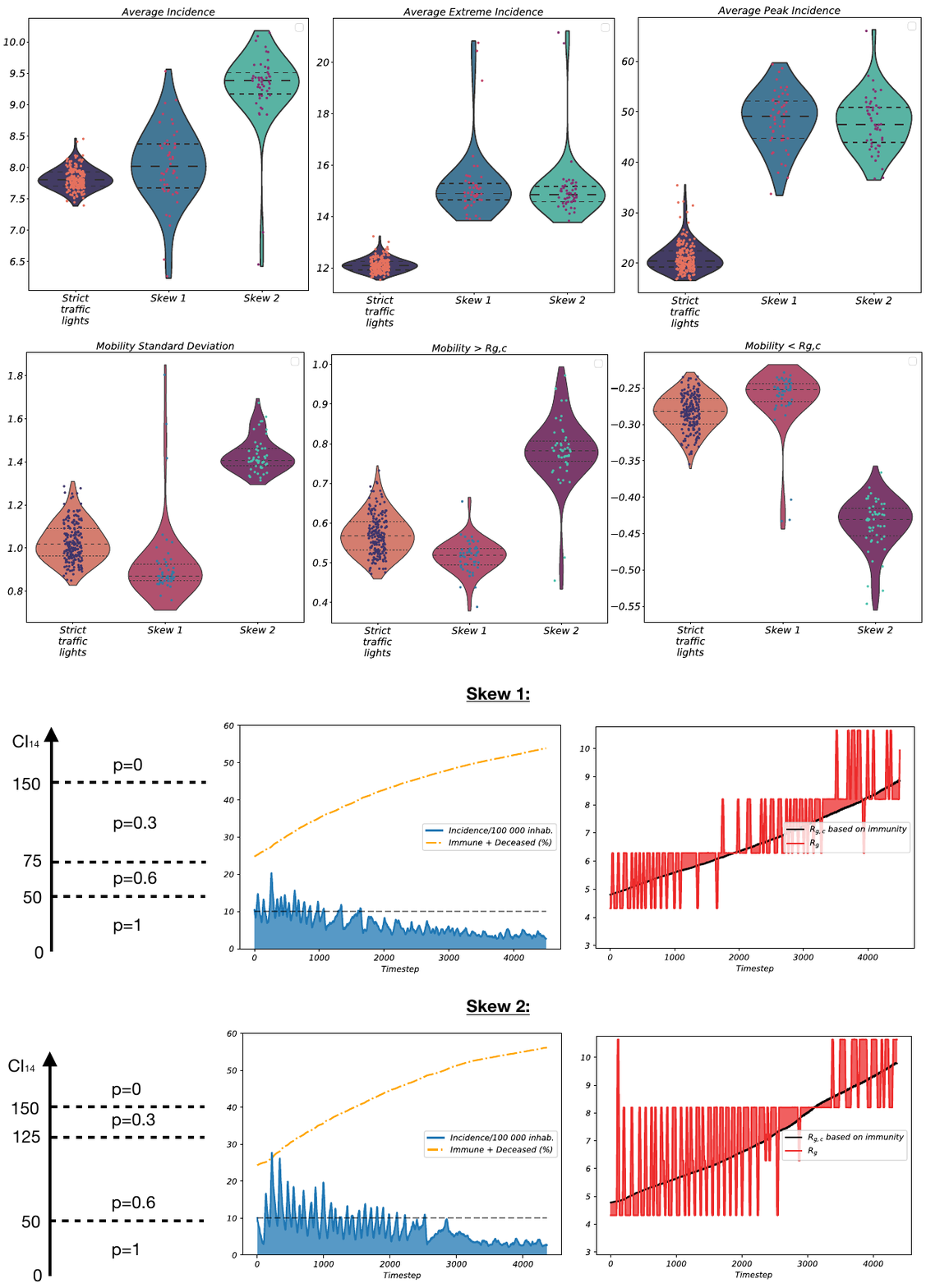}
\caption{Epidemic and mobility examples and statistics corresponding to the strict traffic light intervention described in the main text and a skewing of the $CI_{14}$ thresholds used in our strict traffic light intervention. $T_r = 30$ t.s.}
\label{fig:stats_stab}
\end{figure*}

\begin{figure*}[h!]
\includegraphics[width=17cm,trim=0 100 0 0]{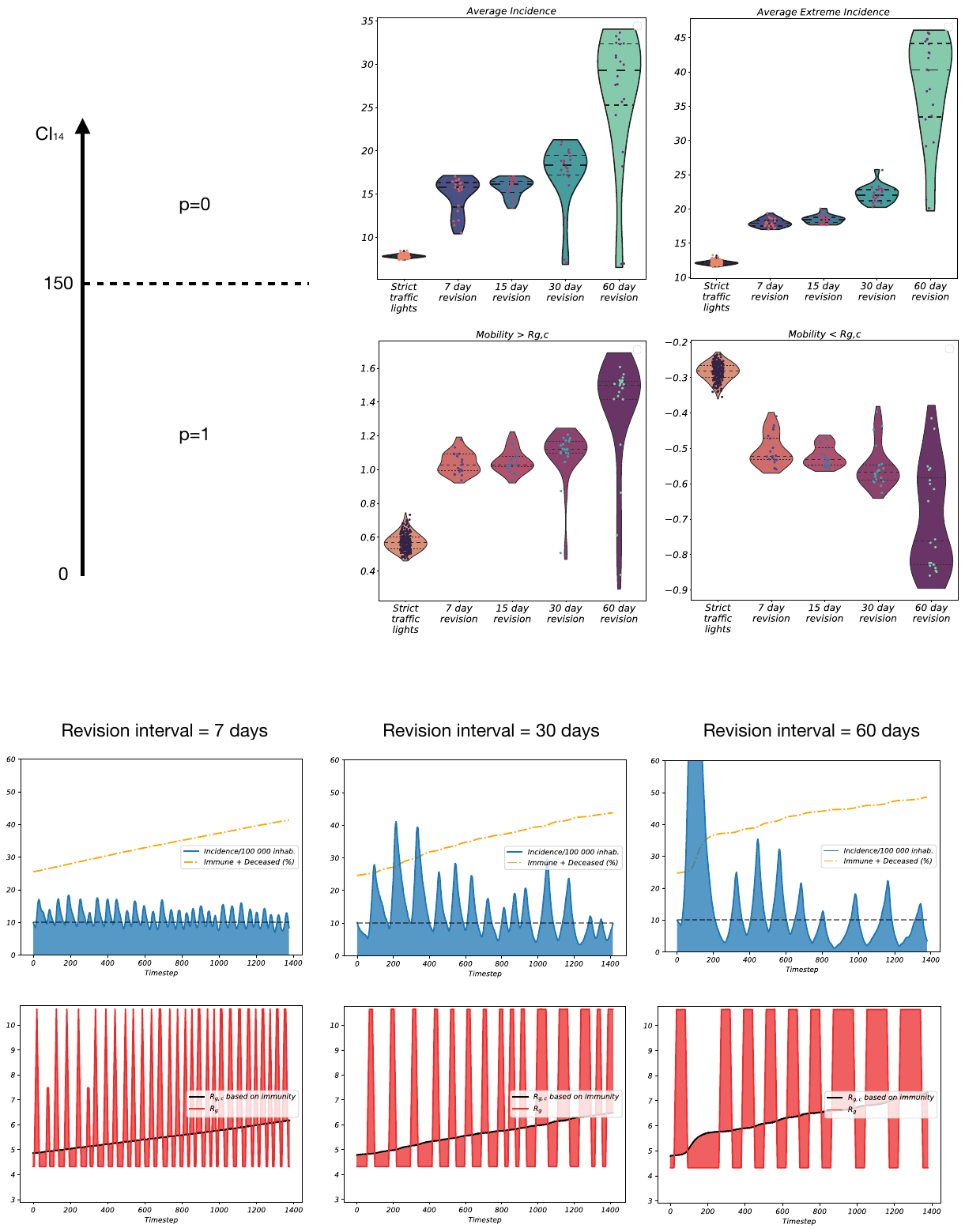}
\caption{Epidemic and mobility examples and statistics corresponding to the strict traffic light intervention described in the main text and a single threshold strict traffic light intervention with variable $T_r$.}
\label{fig:stats_2tf}
\end{figure*}

In our model, the traffic light system has divided the possible mobility spectrum into 4 equally-spaced as a proxy for epidemic containment. Being these values arbitrarily chosen for our model, an exploration of different values can be performed to check the error induced by this proxy. We replicate the strict traffic light intervention, except we now skew the $CI_{14}$ thresholds, making the system default to lower values of $p$ (figure \ref{fig:stats_stab}, skew 1). On the other hand, values can be skewed to default the system to higher values of $p$ (figure \ref{fig:stats_stab}, skew 2). Skewing the values makes it harder for the system to approximate the mobility to it's critical level at each given time and thus incidence metrics result in higher values than before , specially when the system defaults to higher $p$ values (skew 2). Having this in mind, the difference to our proxy isn't big enough to be worried about our assumption of values for the traffic light system. To see the stability of this intervention style drawn to the limit, we can reduce it to 2 mobility states with a single threshold: $p=0$ for $CI_{14} \geq 150$ and $p=1$ for $CI_{14} < 150$. In figure \ref{fig:stats_2tf} we explore how the system reacts to this intervention with different revision intervals $T_r$. With a small enough $T_r$ ($T_r ~ 7-15$ t.s.) the system seems to stabilise into a low oscillatory regime, although not with incidences as low as when using a traffic light system with a finer spectra of states.

\section{Degrees of leniency}

\begin{figure*}[h!]
\includegraphics[width=17cm,trim=0 0 0 0]{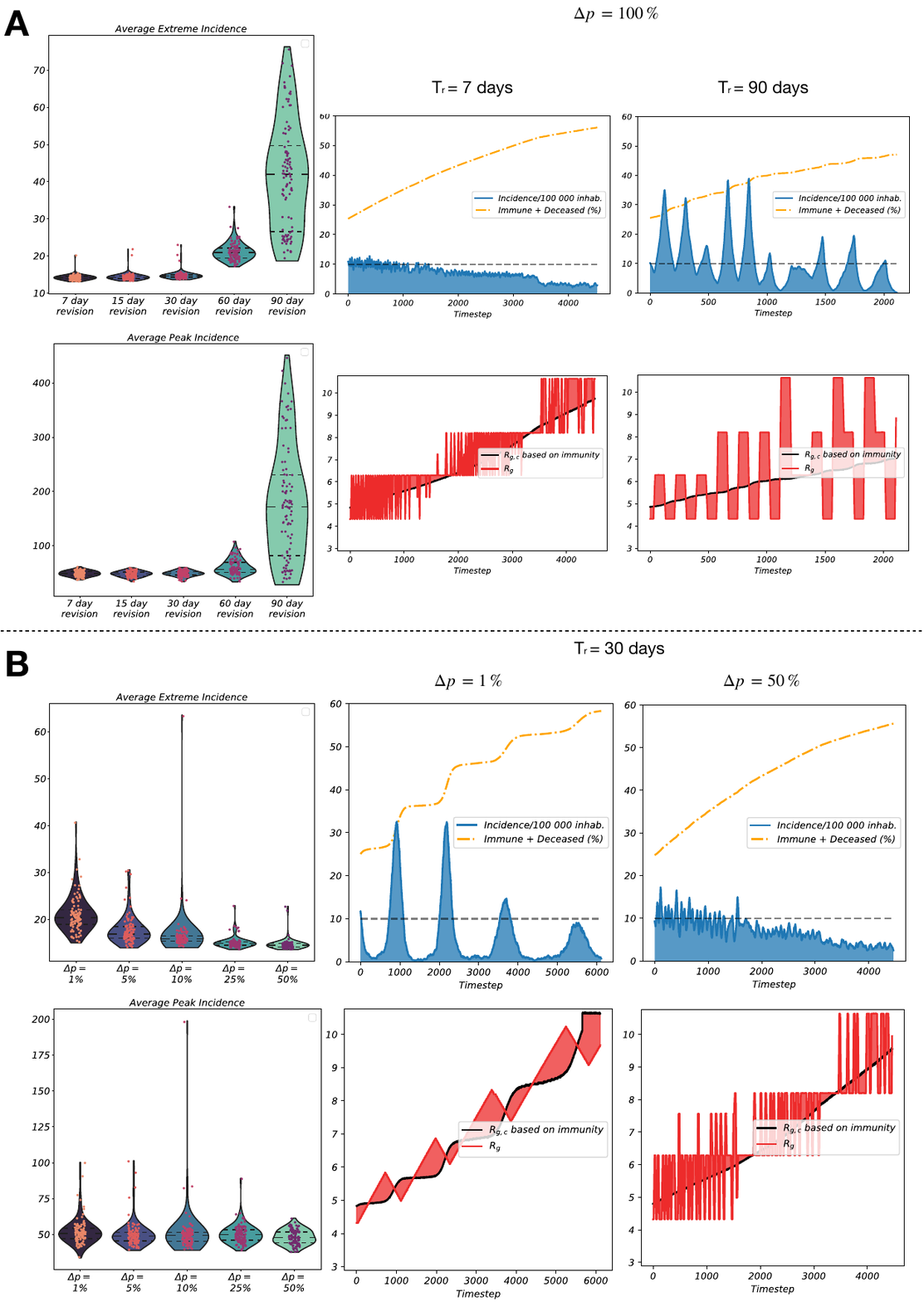}
\caption{A. Extreme incidence and peak incidence for a traffic light system with $\Delta p = 100 \%$ and a variable revision interval $T_r$. B. Extreme incidence and peak incidence for a traffic light system with variable $\Delta p$ and $T_r = 100$}
\label{fig:stats_lenient}
\end{figure*}

When introducing an example of leniency in the application of the traffic light intervention, we chose to revise the mobility every $T_r = 60$ t.s. and only change the mobility parameter $p$ by $\Delta p = 10\%$ when mobility is revised. We know explore both $\Delta p$ and $T_r$ starting from the baseline for strict traffic light interventions, which would be $\Delta p = 100\%$ and $T_r = 30$ t.s.. From figure \ref{fig:stats_lenient}A we can observe how traffic light systems seem to be robust until $T_r > 30$ t.s., at which point the timescale at which contacts, infections and symptoms onset occur is much smaller than the revision interval and the system cannot react in time, producing oscillations and high incidence rates. Figure \ref{fig:stats_lenient}B shows that failing to adjust mobility to the traffic light system by using changes in the parameter smaller than $\Delta p = 0.25$ result in a substantial increase in incidences and oscillations.

\end{document}